\documentclass[sigconf]{acmart}

\def\BibTeX{{\rm B\kern-.05em{\sc i\kern-.025em b}\kern-.08emT\kern-.1667em\lower.7ex\hbox{E}\kern-.125emX}}

\usepackage{amsmath,amssymb,amsfonts}
\usepackage{algorithmic}
\usepackage{graphicx}
\usepackage{textcomp}
\usepackage{xcolor}
\usepackage{caption}
\usepackage{graphicx}
\usepackage{xspace}
\usepackage[ruled]{algorithm2e}
\usepackage{subcaption}
\usepackage{mdframed}
\usepackage{mathtools}

\graphicspath{ {./figs/} }

\newcommand{\quotes}[1]{\textit{``#1''}}
\newcommand{\gni}{\textsc{uGNI}\xspace}
\newcommand{\dmapp}{\textsc{DMAPP}\xspace}

\newcommand{\adaptive}{\textsc{Adaptive}\xspace}
\newcommand{\imb}{\textsc{Increasingly Minimal Bias}\xspace}
\newcommand{\adhigh}{\textsc{Adaptive with High Bias}\xspace}
\newcommand{\default}{\textsc{Default}\xspace}
\newcommand{\awr}{\textsc{Application-Aware}\xspace}
\newcommand{\mpich}{\textsc{MPICH}\xspace}
\newcommand{\pgas}{\textsc{PGAS}\xspace}

\newcommand{\pingpong}{\texttt{ping-pong}\xspace}
\newcommand{\allreduce}{\texttt{allreduce}\xspace}
\newcommand{\alltoall}{\texttt{alltoall}\xspace}
\newcommand{\barrier}{\texttt{barrier}\xspace}
\newcommand{\broadcast}{\texttt{broadcast}\xspace}
\newcommand{\halo}{\texttt{halo3d}\xspace}
\newcommand{\sweep}{\texttt{sweep3d}\xspace}

\newcommand{\milc}{\texttt{MILC}\xspace}
\newcommand{\fft}{\texttt{FFT}\xspace}

\newcommand{\daint}{\textit{Piz Daint}\xspace}
\newcommand{\cori}{\textit{Cori}\xspace}

\newcommand{\review}[1]{{{#1}}}
\newcommand{\htor}[1]{\textcolor{green}{[]}}
\newcommand{\salvo}[1]{\textcolor{purple}{[]}}
\newcommand{\para}[1]{\textcolor{blue}{[]}}
\newcommand{\pages}[1]{\textcolor{red}{[]}}
\newcommand{\todo}[1]{{\color{red} {}}}

\copyrightyear{}
\acmYear{}
\setcopyright{acmlicensed}
\acmConference[]{}{}{}
\acmBooktitle{}
\acmPrice{}
\acmDOI{}
\acmISBN{}

\setcopyright{acmcopyright}
\begin{document}

\copyrightyear{2019} 
\acmYear{2019} 
\acmConference[SC '19]{The International Conference for High Performance Computing, Networking, Storage, and Analysis}{November 17--22, 2019}{Denver, CO, USA}
\acmBooktitle{The International Conference for High Performance Computing, Networking, Storage, and Analysis (SC '19), November 17--22, 2019, Denver, CO, USA}
\acmPrice{15.00}
\acmDOI{10.1145/3295500.3356196}
\acmISBN{978-1-4503-6229-0/19/11}

%
% The "title" command has an optional parameter, allowing the author to define a "short title" to be used in page headers.
\title[Mitigating Network Noise on Dragonfly Networks through Application-Aware Routing]{Mitigating Network Noise on Dragonfly Networks\\through Application-Aware Routing}

%
% The "author" command and its associated commands are used to define the authors and their affiliations.
% Of note is the shared affiliation of the first two authors, and the "authornote" and "authornotemark" commands
% used to denote shared contribution to the research.
\author{Daniele De Sensi}
\email{desensi@di.unipi.it}
\orcid{0000-0002-7244-639X}
\affiliation{%
  \institution{University of Pisa}
  \city{Pisa}
  \country{Italy}
}
\affiliation{%
  \institution{ETH Zurich}
  \city{Zurich}
  \country{Switzerland}
}

\author{Salvatore Di Girolamo}
\email{salvatore.digirolamo@inf.ethz.ch}
%\orcid{XXXX-XXXX-XXXX-XXXX}
\affiliation{%
  \institution{ETH Zurich}
%  \streetaddress{P.O. Box 1212}
  \city{Zurich}
  \country{Switzerland}
}

\author{Torsten Hoefler}
\email{htor@inf.ethz.ch}
%\orcid{XXXX-XXXX-XXXX-XXXX}
\affiliation{%
  \institution{ETH Zurich}
%  \streetaddress{P.O. Box 1212}
  \city{Zurich}
  \country{Switzerland}
}

%
% By default, the full list of authors will be used in the page headers. Often, this list is too long, and will overlap
% other information printed in the page headers. This command allows the author to define a more concise list
% of authors' names for this purpose.
\renewcommand{\shortauthors}{De Sensi et al.}

\begin{abstract}
System noise can negatively impact the performance of HPC systems, and the interconnection network is one of the main factors contributing to this problem.  To mitigate this effect, adaptive routing sends packets on non-minimal paths if they are less congested.  However, while this may mitigate interference caused by congestion, it also generates more traffic since packets traverse additional hops, causing in turn congestion on other applications and on the application itself.
In this paper, we first describe how to estimate network noise.  By following these guidelines, we show how noise can be reduced by using routing algorithms which select minimal paths with a higher probability.  We exploit this knowledge to design an algorithm which changes the probability of selecting minimal paths according to the application characteristics.  We validate our solution on microbenchmarks and real-world applications on two systems relying on a Dragonfly interconnection network, showing noise reduction and performance improvement.
\end{abstract}



%
% Keywords. The author(s) should pick words that accurately describe the work being
% presented. Separate the keywords with commas.

%
% This command processes the author and affiliation and title information and builds
% the first part of the formatted document.
\maketitle

\section{Introduction}
%\pages{1-2}
%\para{the network is the key to large-scale HPC systems}
%\htor{very bad start - boring to study something that has been studied to death}
%The analysis of external effects on application performance has been a well-studied problem, and has been proven that even small delays on a single node can lead to significant performance degradation for large-scale applications, if they interfere with the communication and synchronization between the nodes.

Interconnection networks are the backbone of large-scale supercomputers often connecting tens of thousands of servers. The cost, performance, and maintainability depend on details of the networking technology and topology. Commonly deployed low diameter (e.g., Dragonfly) and hierarchical topologies (e.g., Fat Tree) usually share network resources between applications running in different allocations~\cite{4556717, fattree}. 
This sharing can lead to interference between applications, e.g. if one application communicates heavily and fills shared links and switch buffers with large numbers of packets, another application that only sends small synchronization messages may be severely delayed by the resulting head-of-line blocking.

%\para{network noise}
The caused communication delays can destroy the performance of large-scale applications, similarly to operating system \review{(OS)} noise. \review{Indeed, it has been shown that OS noise limits the scalability of many applications to 15k processes on the 200k core Jaguar supercomputer~\cite{osnoise}. }
Operating systems have been tuned to minimize interference through isolation, for example, by scheduling management tasks on separate cores~\cite{Giampapa:2010:ELS:1884643.1884667}.
However, network isolation and efficient use of computing resources cannot easily be achieved on today's supercomputers. 
The resulting \emph{network noise}  has been observed in various research groups~\cite{sc18, bully, Chunduri:2017:RVX:3126908.3126926}. 
Like operating system noise, we define network noise an \emph{external effect} on application performance, caused by \emph{sharing resources} with activities \emph{outside of the control of the affected application}. Network noise can either be caused by the \review{High-Performance Computing} (HPC) system itself (e.g., control of distributed filesystems) or by other applications running simultaneously (e.g., cross traffic). Thus, in general, \emph{the \review{programmers} of an application cannot avoid noise}, \review{they} can at best mitigate it, for example, using nonblocking collective operations~\cite{hoefler-europvm07,hoefler-sc07}.

%\para{known techniques for network noise avoidance do not work}
At the system level, network noise can be avoided by different application allocations to isolated partitions of the system. For example, to sub-trees in a Fat Tree topology~\cite{fattree:sc18} or groups of a Dragonfly interconnect. However, this is only possible if the allocation sizes exactly match the topology layout and resources are available---introducing such a strategy in any batch system will significantly reduce the system utilization. Thus, network noise is generally accepted in HPC systems and was of not much concern until recently.
Yet, we argue that growing system sizes, as well as the introduction of adaptive routing technologies, aggravate the situation, \review{causing up to 2X slowdowns, as we will show in Section~\ref{sec:evaluation}}. Adaptive routing has been introduced to increase the overall utilization of the network---it is in fact \emph{required} in low-diameter topologies for most traffic patterns~\cite{4556717, slimfly,diameter-2-topos}. The downside of adaptive routing is that even two communicating nodes can congest resources on all paths that the adaptive routing utilizes, not just a single path in static routing. This is often referred to as \emph{packet spraying} and is feared by datacenter operators in so-called \emph{incast} or \emph{hot-spot} (many-to-one) patterns. In addition to this, adaptive routing may be affected by the so-called \textit{phantom congestion} problem~\cite{7056051}. \review{Namely, as congestion information is propagated with some delay, a node may react too late to congestion events.}
In this paper, we will be first to demonstrate and quantify the influence of different routing schemes on network noise in practice.

%\para{analyzing network noise is tricky}
Analyzing network noise in detail is delicate because in practice, when observing application delays, it is hard to distinguish between network noise, operating system noise, and application imbalance.
Other works have used network counters but may run into the fallacy that correlation is not causation by ignoring the aspects of unrelated traffic. 
We will clarify several potential problems in investigations of network noise and develop a set of general guidelines for our analysis.
Using these guidelines, we study the relationship between different adaptive routing schemes, application performance, and network noise. 
Our findings on two large-scale Cray Aries systems, Piz Daint and NERSC's Cori, are \review{remarkable}: not only will changing the adaptive routing mode reduce communication times and speed up applications up to twice, but it will also significantly reduce performance variation. 

We find that the best routing mode that minimizes network noise and maximizes performance depends not only on the characteristics of the allocation but also on the communication load. For example, large-scale alltoall communications are best routed with the default mode \review{whereas} many other communication patterns benefit from mostly minimal routing. 
\review{Because} this all depends on the location of the communication peers, there is no simple static rule to select the best routing mode.
To address this, we develop a simple but effective \review{dynamic} routing library that observes the network state through local network counters and adjusts the routing mode for each message based on application characteristics such as the message size and the observed network state. 
In essence, our routing library \review{is application- and system-aware} and acts as a coarse-grained guide that adjusts the packet-level adaptive routing based on application characteristics and dynamic network state.

In summary, the key contributions of this work are:
\begin{itemize}
    \item We describe mechanisms to gather a detailed understanding of noise in real applications caused by the network using network counters.
    \item We provide a detailed analysis of two real-world systems with low-diameter topologies.
    \item We show that much of the application delay is due to network noise that stems from non-trivial interactions between routing strategies and application characteristics.
    \item We evaluate different adaptive routing strategies using bias on Cray Aries (Cascade) systems.
    \item We develop an application-aware routing library that routes different applications and application phases with different routing modes, leading to speedups of 2x on some microbenchmarks and real applications.
\end{itemize}

\section{Network performance counters on Cray Aries}\label{sec:background}
We describe in this section the topology of the Cray Aries Dragonfly network, the different routing algorithms available, and the network counters provided. Eventually, we introduce a performance model which, by using selected network counters, can be used to estimate the transmission time of a message.

\subsection{The Cray Aries Interconnect}

\begin{figure}[htpb]
    \centering
    \includegraphics[width=\columnwidth]{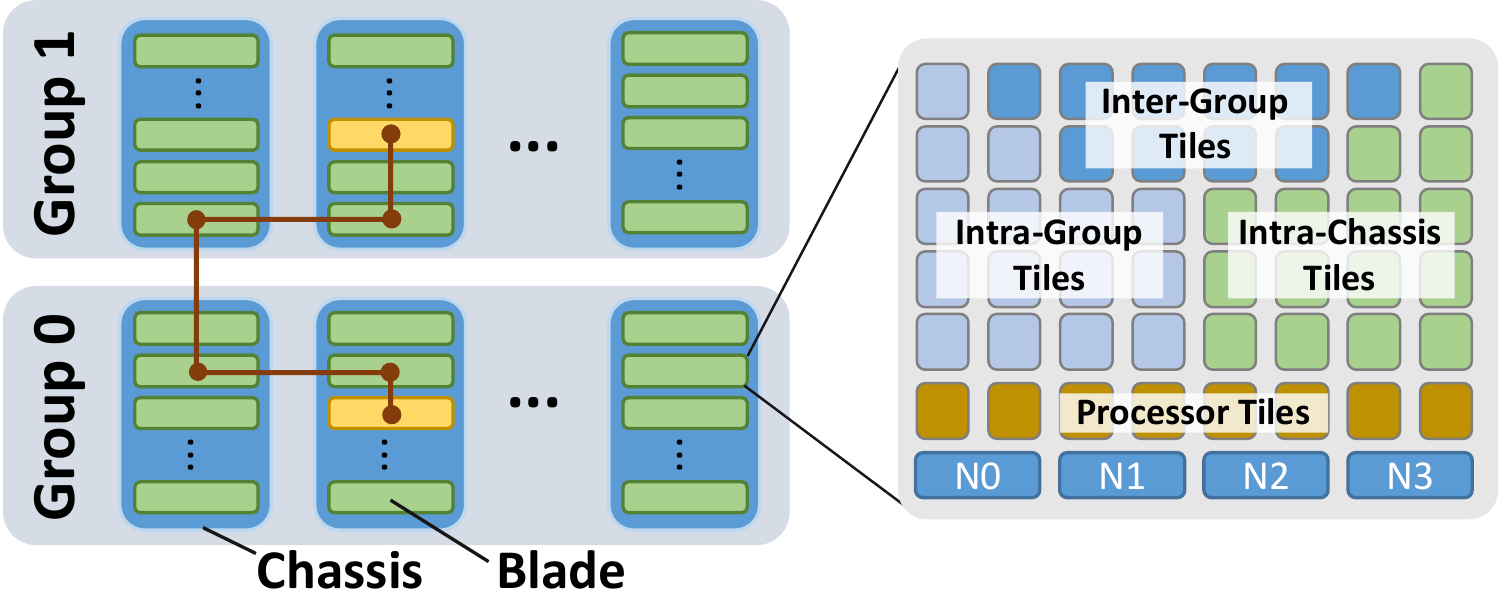}
    \caption{Cray Aries Topology and Aries Device.}
    \label{fig:aries}
\end{figure}

The Cray Aries Network~\cite{crayxcpdf} is a high performance interconnect based on
the Dragonfly topology~\cite{4556717}. The underlying idea of the Dragonfly
topology is to partition compute nodes and routers in fully connected groups
that act as very high radix virtual routers. 
%
% Topology
The Aries interconnect is organized in three connectivity tiers: groups, chassis, and blades.
Each group contains six
chassis and within each chassis there are sixteen blades. 
%A group is distributed over two cabinets. 
Each blade contains the Aries router and four nodes.
%
% Connectivity
Figure ~\ref{fig:aries} sketches two groups of an Aries interconnect and the
internal organization of an Aries device. An Aries device is a system-on-chip comprising
a 48-port router and 4 \review{Network Interface Controllers} (NICs). The router is organized in tiles: each
tile provides a bidirectional link with a bandwidth ranging from 4.7 and 5.25 GB/s per
direction (depending
on whether the link is optical or electrical). 
Each router can have up to ten optical connections to routers in different 
groups: the total number of groups is constrained by the number of routers
per group. Often, systems are configured to use more than one tile per inter-group
connection, increasing the inter-group bandwidth. 
The router is connected to all the other routers on the same chassis using
the 15 intra-chassis tiles and to 5 other routers sitting on
different chassis of the same group using three \review{intra-group} tiles per connection.

\review{As a consequence, routers inside a group are not fully connected. For example, in Figure~\ref{fig:aries}, any router in Group 1 is directly connected to all the routers in the same \quotes{column} (i.e., in the same chassis), and to all the routers in the same \quotes{row} (i.e., all the routers in the same position but on the other chassis).}
\review{For this reason,} to reach its destination, a packet traverses between one and six tiles of the Aries
interconnect. Figure~\ref{fig:aries} shows the example of a 5-hop minimal path
between the two yellow blades, in the case where there is no direct link between their \textit{inter-group} tiles. 
\review{In this example, we assume that Group 0 is connected to Group 1 through the second blade in its first chassis. When a node on the yellow blade on Group 0 sends a packet to a node on the yellow blade on Group 1, the packet needs first to be routed to the second blade on its chassis (since it is the only blade directly connected to the second blade on the first chassis). Multiple paths are available between any pair of blades and the routing algorithm could take different decisions (e.g., routing first the packet from the yellow blade to the third blade in the first chassis and then do an intra-chassis hop on the first chassis).} 
The tiles have a finite input queue and use
a credit flow control scheme to avoid overflowing 
the destination tile. As a
consequence, a packet may stall in any of the tiles that are on its path 
to the destination. 

The router also has 8 processor tiles to connect to the 4 Aries 
NICs included on the Aries Device (N0, \ldots, N3 in the figure) 
and each pair of NICs shares four processors tiles. 
The NICs are connected to the respective
compute nodes via independent x16 PCIe Gen3 host interfaces.
The compute nodes can issue commands to the NIC via the host interface: in case of a
data movement command, the NIC is in charge to packetize the data and issue up
to 64 bytes request packet to the connected processor tiles. Each request packet is then acknowledged by a response packet.

Data can be carried either in request packets (in case of \textit{RDMA PUT} calls), or in response packets (for \textit{RDMA GET} calls). \review{Each packet is composed by a number of \textit{flits}.} When a request is sent, the NIC splits the request packet in one \textit{header} flit plus one to four payload flits\footnote{The size of the flit may be different in other points in the network (e.g., on network tiles), thus changing the number of flits which make up a packet.}, and transmits one flit per clock cycle. It is thus possible to estimate the number of packets and flits which will be sent by the NIC by knowing the type of \textit{RDMA} call and the size of the application message (i.e., we will have 1 packet every 64 bytes, made of 1 request flit for \textit{GET}s and 5 request flits for \textit{PUT}s).

In this paper we will consider two different machines based on Cray Aries Network:
\begin{description}
\item[\daint] A Cray XC50 system hosted by CSCS, with compute nodes equipped with a 12-core Intel Xeon E5-2690 \review{v3} CPU with 64 GiB RAM \review{and with Hyper-Threading support}. 
\item[\cori] A Cray XC40 system hosted by NERSC, with compute nodes equipped with a 16-core Intel Xeon E5-2698 \review{v3} CPU with 128 GiB RAM  \review{and with Hyper-Threading support}.
\end{description}

%- topology picture?
%- describe the topology and the location of counters. 
%- The connectivity is defined by the router
%- groups fully connected, nodes in each group not fully connected. Given two nodes,  length of minimal paths may vary 
%- Introduce daint/cori

%A router is shared between four nodes, that are interfaced through 8 processor tiles
%- the tile business and whatnot

%- each router has X tiles and is shared by 4 nodes. 2 processors tiles shared by 2 nodes. One NIC per node?

\subsection{Adaptive routing and bias}\label{sec:background:routing}
On the Aries network, each packet can be independently routed, and \review{because} two nodes can be connected by several (minimal and non-minimal) paths, adaptive routing is used so that packets are sent to less congested paths. The adaptive routing algorithm adopted is a variation of UGAL routing~\cite{ugal}. Packets sent on non-minimal paths will traverse an intermediate group connected to both source and destination groups, increasing the maximum number of hops up to 10 on the largest networks~\cite{crayxcpdf}. Every time a packet is sent, two minimal and two non-minimal paths are randomly selected, and the congestion of these paths is estimated by using both local information (e.g., queue occupancy) and estimation of \textit{far-end} congestion, based on the current flow credits available. 

However, due to the long inter-router latency, credit information may be delayed, resulting in inaccurate congestion information, leading the adaptive algorithm to select non-minimal paths even if that was not necessary anymore~\cite{7056051}. 
For this reason, known as \textit{phantom congestion}, a \textit{bias} value can be added to the congestion estimated for non-minimal paths, so that the higher is the bias, the higher is the probability that the packet will be routed on a minimal path. Although it is not possible to set an arbitrary value for the bias, for MPI applications, the bias can be selected by the user among a restricted set of possibilities (which exact value is not public) by setting the \texttt{MPICH\_GNI\_ROUTING\_MODE} environment variable before starting the application\footnote{The routing algorithm for \texttt{MPI\_Alltoall} calls can be separately selected through the \texttt{MPICH\_GNI\_A2A\_ROUTING\_MODE} environment variable.}. 

This variable can be set to one of the following values:

\begin{description}
\item[ADAPTIVE\_0] No bias is added. We will refer to this algorithm as \adaptive.
\item[ADAPTIVE\_1] Bias towards minimal routing increases as the packet approaches the destination~\cite{imb}. It is the default routing algorithm used for \texttt{MPI\_Alltoall} communications. We will refer to this algorithm as \imb.
\item[ADAPTIVE\_2] A \textit{low} bias is added. 
\item[ADAPTIVE\_3] \review{A} \textit{high} bias is added. We will refer to this algorithm as \adhigh.
\end{description}

Moreover, this variable can also be used to enforce deterministic routing rather than adaptive routing, by setting one of the following values:
\begin{description}
\item[MIN\_HASH] Packets are always routed minimally, and the path is selected based on a hash of some fields of the packet header. 
\item[NMIN\_HASH] Packets are always routed non-minimally, and the path is selected based on a hash of some fields of the packet header.
\item[IN\_ORDER] Packets are always routed minimally, and the packets are received in the same order they were transmitted.
\end{description}

In this work, we will only focus on ADAPTIVE\_0, ADAPTIVE\_1, and ADAPTIVE\_3 routing algorithms. Indeed, the performance of ADAPTIVE\_2 lie between those of ADAPTIVE\_0 and ADAPTIVE\_3 \review{because} its bias lies between that of these two algorithms. Moreover, we will not consider MIN\_HASH, NMIN\_HASH, and IN\_ORDER \review{because} they are not adaptive algorithms.
%We observed for \minhash and \inorder have similar or worst performance than \adhigh since they are deterministic routing algorithms and do not react to congestion. For the same reason, and also because it never sends packets on minimal paths, \nminhash performs worst than the \adaptive routing algorithm. 
%- virtual channels needed  (not sure this is a relevant information for this work, maybe just mention it without going too much into details)

%- How to change the routing with environment variable

\subsection{Counted events}\label{sec:background:counters}
Aries provides several network counters, which can either be accessed either by using the PAPI library~\cite{papi} or the CrayPat tool, allowing the user to  monitor the impact of the network on his application and vice-versa. Counters are present on both NICs and processor/network tiles. However, users can only access counters on the NICs and tiles associated with their jobs. \review{Because} adaptive routing is used, packets may traverse routers which are entirely allocated to other jobs, and by relying on network tiles counters, we would only have a partial view of the impact of our application traffic on the network. Moreover, each tile can be traversed by traffic coming for different jobs, and there is no way to isolate the contribution of each individual job. 

\review{For these reasons, we rely solely on the NIC network counters. Among the different provided counters, we will focus on the following ones, which provide enough information for our purposes:}
\begin{description}
\item[Request Flits] Number of request flits sent.
\item[Request Flits Stalled Cycles] This counter increments every clock cycle a ready-to-forward flit is not forwarded because of back-pressure. The ratio between this counter and the number of request flits represents the average number of cycles a flit must wait before being transmitted.
\item[Request Packets] Number of request packets sent.
\item[Request Packets Cumulative Latency] Cumulative latency (microseconds) across all the request-response packets pairs. By dividing this counter for the previous one we get the average packet latency. \review{This counter does not include the time a flit waits in NIC queues before being transmitted.}
\end{description}

\review{Detailed information about all the network counters available on Aries can be found on Cray's technical documentation~\cite{arieshwcounters}.}

\subsection{Performance model}\label{sec:background:model}
We now show how to model how the significant network counters described in Section~\ref{sec:background:counters} influence application performance.
Inspired by the well-known LogP model~\cite{Culler:1996:LPM:240455.240477}, we develop a model including the average stall cycles and the latency that we observe through the counters.
We define with $L$ the packet latency in NIC cycles, with $RTT$ the round-trip-time of a flit, with $s$ the average number of cycles a flit waits (due to stalls) before being transmitted, with $k$ the number of flits per packet, and with $f$ the number of flits which compose the application message.

\begin{figure}[htpb]
    \centering
    \includegraphics[width=\columnwidth]{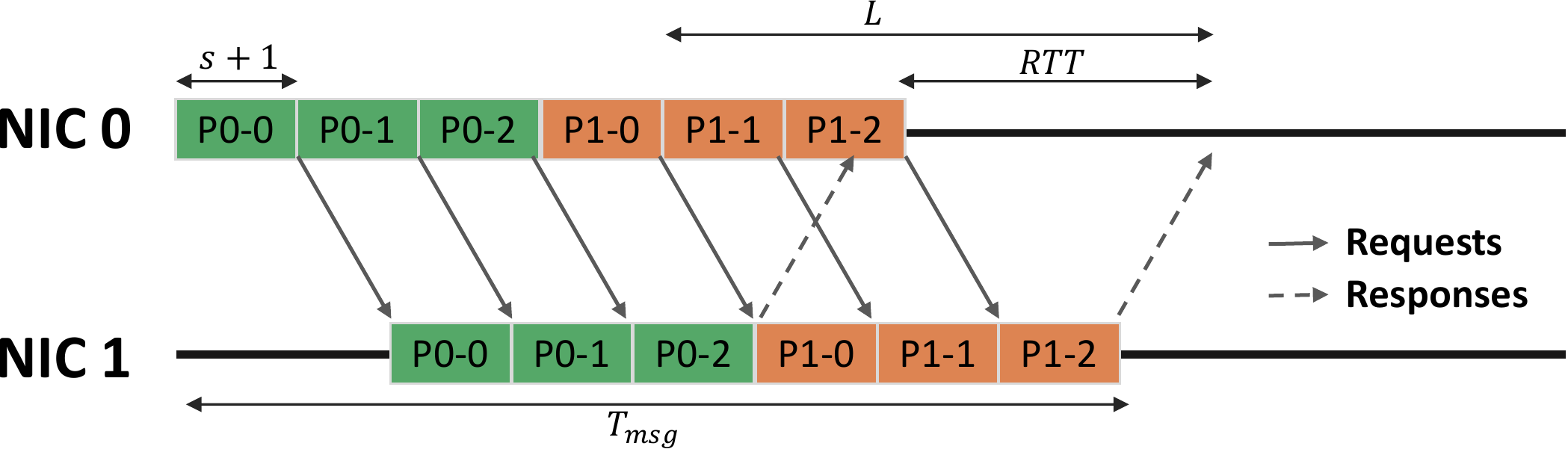}
    \caption{Relationship among NIC performance counters and transmission time of an application message.}
    \label{fig:background:model}
\end{figure}

We illustrate all parameters in Figure~\ref{fig:background:model} for a \textit{PUT} message and note that a similar model can be derived for \textit{GET} messages. In the picture we show a scenario where the application message is decomposed in six flits ($f$), divided into two network packets \review{(shown in green (P0) and orange (P1))}. Each network packet \review{comprises} three request flits ($k$) traveling from the sender's NIC to the receiver's NIC, and one response flit acknowledging the reception of the request and traveling in the opposite direction. \review{The way in which we defined $s$ and $L$ reflects the quantities measured by the counters available on Aries Networks (described in Section~\ref{sec:background:counters}). For example, $L$ measures the latency between the sending of the first request flit of the packet (excluding the waiting time $s+1$) and reception of the last response flit\footnote{The counter on Aries NICs provides a measure of the latency in microseconds. This can be easily converted to NIC cycles given the clock frequency of the NIC.}.}

First of all, for reasons which will become clearer in Section~\ref{sec:estimation:contention}, we are interested in estimating the impact of the network without including any host-side delays. This can be done by measuring $T_{msg}$, i.e., how many cycles elapse between the reception of a PUT call by the sender's NIC to the moment when the last flit is delivered to the receiver's NIC. 
In the absence of stalls, the NIC can transmit one flit per clock cycle. For this reason, if stalls are present, \review{each flit is stalled on average for $s$ cycles and the NIC transmits one flit every $s + 1$ cycles.}

We can then define the transmission time of a packet as the time the first flit takes to reach the receiver's NIC (i.e., $\frac{RTT}{2}$), plus the time required to transmit $f$ flits. The round-trip-time $RTT$ can be obtained by removing from the latency $L$ the time the sender's NIC spent in transmitting $k - 1$ flits, i.e., $RTT = L - (k - 1) \cdot (s + 1)$. However, $s$ is usually some order of magnitude smaller\footnote{For readability reasons, proportions in Figure~\ref{fig:background:model} do not represent the true scales of $L$ and $s$, i.e. $s$ is in the order of a few cycles and $L$ is in the order of thousands of cycles.} than $L$ and we can approximate ${RTT}$ as ${L}$, thus obtaining:

\begin{equation}\label{eq:tmsg_2}
T_{msg} = \frac{L}{2} + f \cdot (s + 1)
\end{equation}

However, Aries NICs can have at most 1024 outstanding packets. For this reason, if more than 1024 packets need to be sent, the NIC must wait for the reception of the responses for sent packets before transmitting additional packets. As a consequence, the transmission of some packets may be not fully overlapped and this would increase $T_{msg}$ by a quantity proportional to the latency. In the best-case scenario, this would happen only once every 1024 packets. Defining $p$ as the number of the packets, the transmission time of the message would then be:

\begin{equation}\label{eq:tmsg}
\begin{split}
T_{msg} &\approxeq {\frac{p}{1024}} \cdot L + \frac{L}{2} + f \cdot (s + 1) = \\ &= {\frac{p + 512}{1024}} \cdot L + f \cdot (s + 1)
\end{split}
\end{equation}

\review{Whereas} $L$ and $s$ can be obtained through network counters, $f$ and $p$ can be estimated  from the message size and the type of \textit{RDMA} request issued. To validate this model, we compared the estimations made by the model with the actual execution time of a \pingpong benchmark executed over 40 different allocations on the \daint machine, obtaining an average $79\%$ correlation on different message sizes, ranging from 128 bytes to \review{16MiB}.
We will leverage this performance model in Section~\ref{sec:awr} to analyze the impact of the routing algorithm on the network noise and to design our application-aware routing algorithm.

\section{Network Noise Estimation}\label{sec:estimation}
%\pages{1-2}
Estimating the true impact of network noise on an application is a complex task. 
The main problem is to isolate the effects of network noise from other effects causing performance variability such as (operating) system noise or varying resource mapping strategies. 
In the following, we will describe and categorize such effects and derive simple but important rules for designing experiments with network noise. 
%
%We note that these strategies for experimental design have not been described in previous works and thus mislead conclusions based on the reported data.
We note that these strategies for experimental design have not been described in previous works and, \review{as we will show in this section, may lead to overestimation of network noise}. 
We also quantify the potential influence that each strategy may have on the final outcome if it was ignored.
\review{Although in the following we will apply these rules to analyze noise on Dragonfly networks, they are general enough to be applied to other interconnection networks as well.} 

\subsection{Fixing the allocation}\label{sec:estimation:allocation}

Process-to-node allocation strategies attempt to solve a complex scheduling problem by trying to find the best balance between fairness, time to completion, topology mapping, throughput, and many other metrics. Thus, it is not rare that the processes of a particular compute job are scattered throughout the network---
%creating an attack-surface for network noise.
\review{making it vulnerable to network noise}.
Specifically in low-diameter networks such as Dragonfly~\cite{4556717, diam2paper}, paths between two arbitrary nodes often have widely different performance. Thus, changing the allocation will change the performance, even in the absence of any network noise.

\begin{figure}[htpb]
    \centering
    \includegraphics[width=\columnwidth]{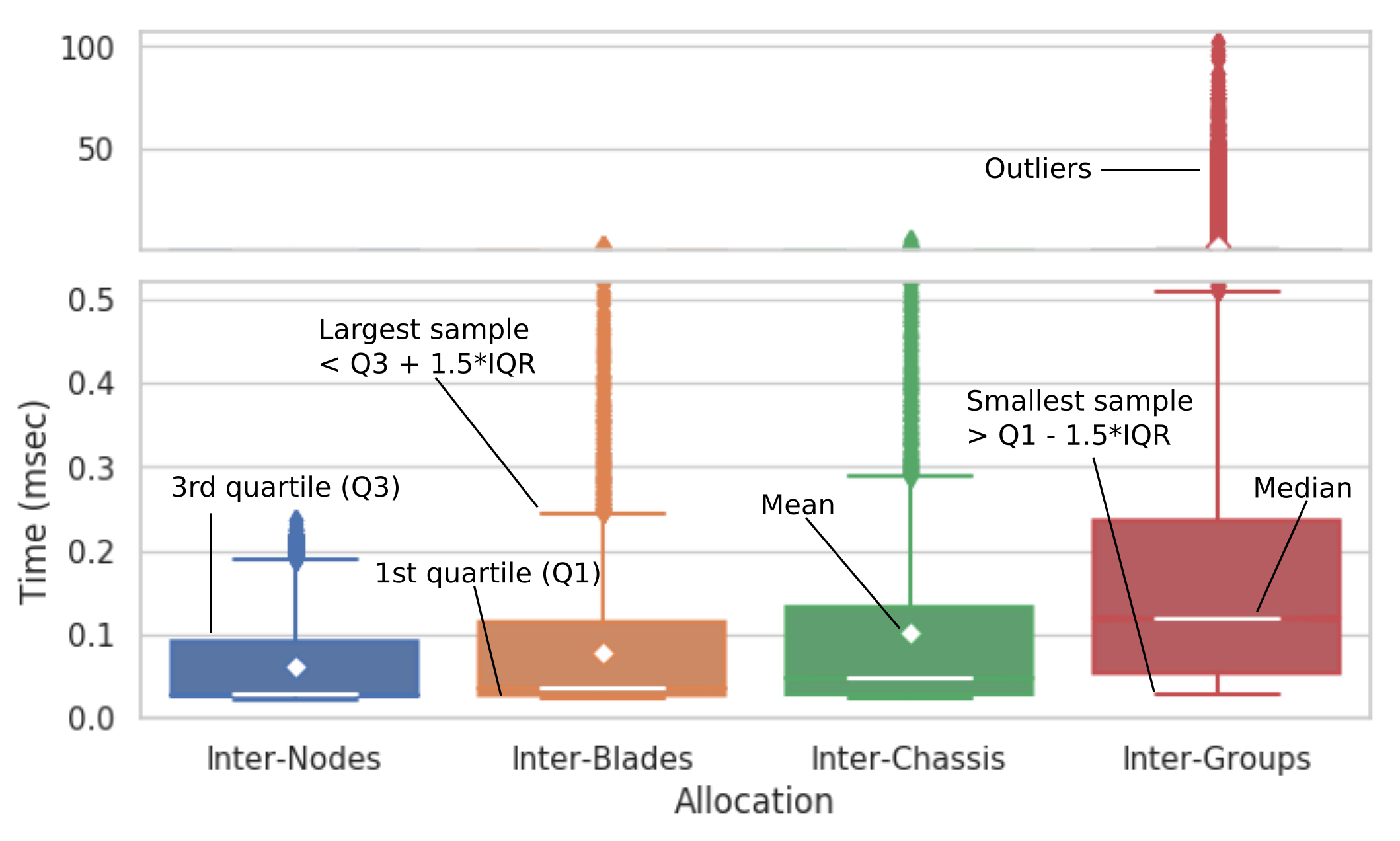}
    \caption{Performance of the \pingpong benchmark for  different allocations on the \daint system. IQR = Inter-Quartile Range (i.e. $Q3 - Q1$).}
    \label{fig:estimation:allocation}
\end{figure}

Figure~\ref{fig:estimation:allocation} quantifies the impact of different allocations on a simple ping-pong benchmark with a 16\review{KiB} message between two nodes on the \daint system. We compare allocations with varying process-to-node mappings for source and destination process: the same blade (\textit{Inter-Nodes}), two different blades (\textit{Inter-Blades}), two nodes on different chassis (\textit{Inter-Chassis}), and two nodes on different groups (\textit{Inter-Groups}). Each of these tests has been run for 5 hours recording one round-trip per second in the same allocation. It illustrates how different allocation strategies not only change the mean but also the variance and distribution of measured performance values. The 95\% confidence interval for the median is represented as a notch around the median (not visible on this specific plot \review{because} its width is $< 5\%$ of the median).

The figure clearly shows not only a growing median time but a massively increasing variance. Some outliers in the inter-group allocation are three orders of magnitude larger than the median, which even pulls the average into the regime of outliers for inter-group communication! Furthermore, \review{whereas} the inter-chassis median is barely higher than the inter-nodes median, the numerous outliers increase the average of the former to be nearly twice as high. We gathered similar results for more complex collective communication benchmarks. This demonstrates the huge potential influence (three orders of magnitude!) that different allocations could have when collecting data about network noise.

\emph{To isolate the effects of network noise, we must avoid effects from varying processor allocations. The simplest strategy is to only compare and analyze results that were collected within the same allocation/job execution.}

\subsection{Correlation is not (always) causation}\label{sec:estimation:correlation}
A common approach in analyzing the impact of network noise is to correlate the execution time of network-intensive applications to the network traffic intensity, measured through network tiles counters~\cite{allreduce}. For example, let us assume we observed an increase in both the execution time and the number of flits that traversed the routers used by the application. At a first sight, we may conclude that \review{because} there was more traffic, each packet had to wait for a longer time before traversing a link, increasing the message latency and slowing down the entire application. However, if the application was delayed for reasons not related to network (e.g., OS noise, imbalance between nodes, etc...), we would observe the network for a longer period and we would generally see a higher number of flits, due to other applications sending packets through the routers we are monitoring. 

% 16 nodes, 5 blades
\begin{table}[htbp]
  \begin{tabular}{lcc}
    \toprule
    (Idle) Time (sec) & Incoming Flits & Stalled Cycles \\
    \midrule
    1 & 110M & 94M  \\ 
    2 & 255M & 157M \\ 
    \bottomrule
  \end{tabular}
    \caption{Relation of (Idle) Time, Flits, and Stalls}
  \label{tab:correlation}
\end{table}

Table~\ref{tab:correlation} demonstrates the effect. It shows the (idle) execution time, the number of flits and the number of stalled cycles for an application executed on 16 nodes, spanning over 5 blades. The application just waits 1 or 2 seconds, respectively, and then terminates. \review{Although the execution time and the number of flits are correlated}, it is clear that the longer execution time caused an increase in the number of observed flits, rather than the other way around.

This is also a relevant problem for those solutions that try to correlate network counters to execution time using machine learning approaches~\cite{sc18}. Indeed, such algorithms may conclude that the network intensity is the most relevant feature having an impact on the execution time even if there is no causal relationship.
\emph{This issue can be mitigated by normalizing the counters with respect to the observation interval or can be completely avoided by relying on NICs counters measuring latency and stalls, \review{because} they have a direct effect on the application performance}.

\subsection{Communication time variation is not network noise}\label{sec:estimation:contention}
\begin{figure}[htpb]
    \centering
    \includegraphics[width=\columnwidth]{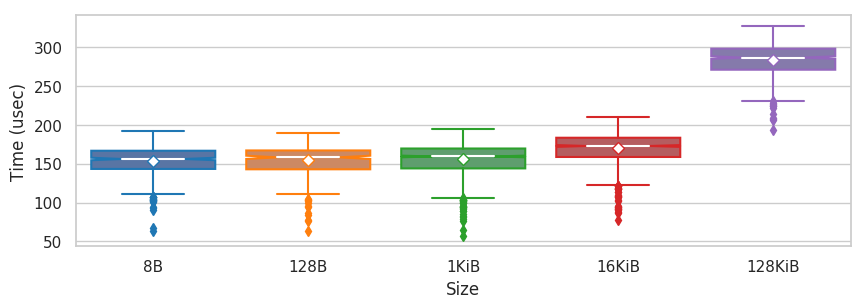}
    \caption{Execution time distribution of an \texttt{MPI\_Alltoall} collective operation executed by 8 processes running on the same node of the \daint machine, for different message sizes.}
    \label{fig:estimation:contention}
\end{figure}

Another common way to estimate network noise is to analyze the variability in the execution time of the communication phases of the application, for example by focusing on the execution time of MPI routines~\cite{sc18, allreduce, Chunduri:2017:RVX:3126908.3126926}. 
However, especially for collective operations, this would also include other delays which do not depend on the network, such as OS noise~\cite{osnoise}, synchronization overheads due to application imbalance~\cite{imbalance}, or contention for shared resources. 
To provide evidence that not all the variations we observe on the execution time of the network routines are caused by network noise, we show in Figure~\ref{fig:estimation:contention} the performance of an \texttt{MPI\_Alltoall} collective operation executed by 8 processes running on \textbf{the same node} of the \daint machine, for different message sizes. Even though the network is not used at all, we can observe a significant  performance variability. This demonstrates that varying execution times of communication operations are not always a good indicator of network noise. 

\begin{figure}[htpb]
    \centering
    \includegraphics[width=\columnwidth]{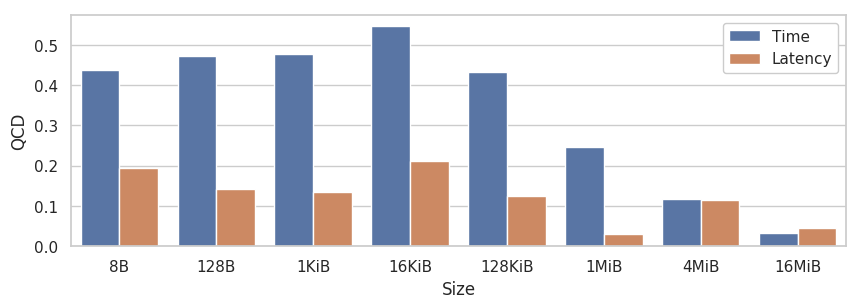}
    \caption{Quartile coefficient of dispersion of execution time and packet latencies, for a \pingpong benchmark between two different groups on the \daint machine, for different message sizes.}
    \label{fig:estimation:latencyvariability}
\end{figure}

Moreover, this problem is not only relevant when multiple concurrent processes per node are used. Indeed, Figure~\ref{fig:estimation:latencyvariability} reports the variability of both the execution time and the network packet latency of a the \pingpong benchmark between two nodes in two different groups on the \daint machine, with only one process per node.
%binded to the second core to reduce the impact of OS noise~\cite{}\todo{There was a paper claiming that but I don't remember by whom}. 
\review{Because} in this specific case the output requests did not experience any stall, latency provides a good approximation of the variability in the time required to transmit the message.

We measure the variability by using the \textit{Quartile Coefficient of Dispersion} (QCD), defined as:
\[
QCD = \frac{Q3 - Q1}{Q3 + Q1}
\]
where $Q3$ and $Q1$ are the third and the first quartile respectively. This would give us a measure of how much the data is concentrated around the median, i.e., the higher the value, the higher is the variability in the data.

As we can see from the figure, even when removing node-side contention, the variability in the execution time of communication routines is still an overestimation of the network noise. This is particularly true for small messages, \review{whereas} the impact of latency on execution time decreases for larger messages.

\emph{To avoid this problem, we must only consider the delays which are induced by the network, for example by using network counters that measure the actual packet latencies, and which do not include the host-side delays. } 

%Other approaches estimate network noise by considering other metrics such as the number of flits or the number of stalled cycles observed on the network tiles. Besides being affected by the issue we described in Section~\ref{sec:estimation:correlation}, this solution also has two other problems. The first one is that, in general, it is only possible to gather statistics for the routers allocated to the application, but since the packet may also traverse routers which are not allocated by the application, by doing so we would only have a partial view of the problem. This issue is not present when considering packet latencies, since we would account for the entire journey of the packet from the source to the destination. The second problem is that, since packets from different applications traverse a given router, not all the stalls we observe affect our application, thus leading to an overestimation of the network noise.

% Paper: Diagnosing the Causes and Severity of One-Sided Message Contention analyzes the impact of contention when changing the number of nodes (not the number of cores)
% Paper: MPI+Threads: Runtime Contention and Remedies shows the impact of contention when varying the number of threads in MPI applications (part of this contention is due to threading and not to access to the NIC)
% (NOTE: \textit{The GNI Provider Layer for OFI libfabric} shows some plot with NIC contention effects. Double check? We could try to get some data by our own)

\section{Routing Impact on Network Noise}\label{sec:awr}
%\pages{1-2}
After establishing a baseline for measuring network noise in isolation, we now analyze the impact of the routing algorithm on noise. We will show how a significant share of the network noise on a Dragonfly network is caused by the adaptive routing algorithm. We analyze the causes of this behavior and we will exploit this information to design an algorithm that, at runtime, can detect and mitigate network noise and improve application performance. \review{Because the algorithm does not make any assumption on the network topology, it could also be used to mitigate network noise on other networks relying on non-minimal adaptive routing.}

\subsection{Interactions between noise and routing}
As described in Section~\ref{sec:background:routing}, each time a packet is sent, the adaptive routing algorithm will estimate the congestion of two minimal and two non-minimal paths (chosen randomly), and will then route the packet on the path which is estimated to be the least congested one. 
This process introduces variability in the packets latencies for different reasons. First of all, each packet takes a different path, with a different number of hops and thus a different latency. Moreover, if on one side by taking non-minimal paths the packets avoid congestion, by traversing more routers there will be more traffic on the network, generating congestion on other applications but also on the application itself.
%In general, the traffic generated by sending all the packets non-minimally is up to twice the traffic generated when all the packets are sent on minimal paths~\cite{}.
\begin{figure}[htpb]
    \centering
    \includegraphics[width=0.8\columnwidth]{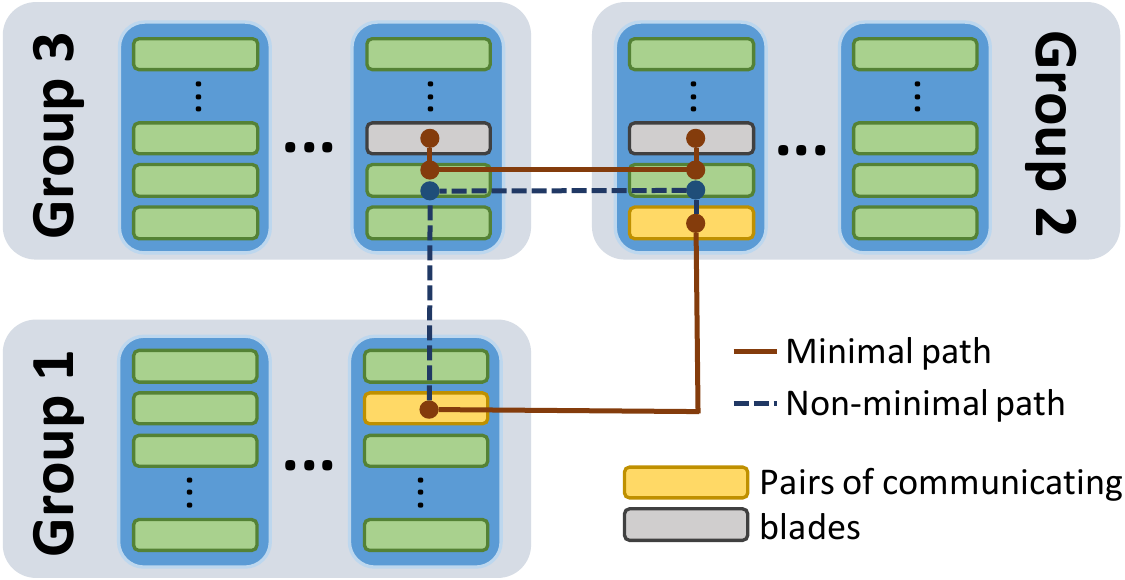}
    \caption{Example of network noise generated by adaptive routing.}
    \label{fig:mpcong}
\end{figure}

For example, let us consider the scenario depicted in Figure~\ref{fig:mpcong}, where the node on the yellow blade on group 1 needs to send a message to the  node on the yellow blade on group 2. After estimating the congestion on both the minimal and non-minimal paths, the node decides to send the packet on the non-minimal path traversing group 3. However, in the meanwhile, the two nodes on the gray blades start to communicate and the traffic generated by the yellow blade would introduce noise on the application running on the gray blades. If the two gray blades were allocated to the same job running on the yellow blades, this would introduce noise on the application itself. Clearly, a similar situation could happen also if only minimal paths are selected. However, due to the higher number of hops, the problem is more severe when using non-minimal paths.

\begin{figure*}[htbp]
\begin{center}
\begin{subfigure}[b]{.18\textwidth}
    \centering
    \includegraphics[width=\columnwidth]{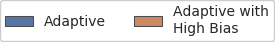}
\end{subfigure}%
\end{center}

\begin{subfigure}[b]{.25\textwidth}
    \centering
    \includegraphics[width=\columnwidth]{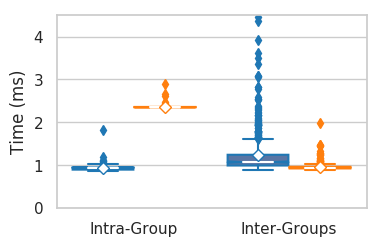}
    \caption{Execution time.}
    \label{fig:awr:routing:time}
\end{subfigure}%
\hfill
\begin{subfigure}[b]{.25\textwidth}
    \centering
    \includegraphics[width=\columnwidth]{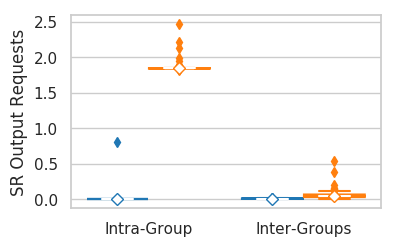}
    \caption{Stalls ratio $s$.}
    \label{fig:awr:routing:srout}
\end{subfigure}%
\hfill
\begin{subfigure}[b]{.25\textwidth}
    \centering
    \includegraphics[width=\columnwidth]{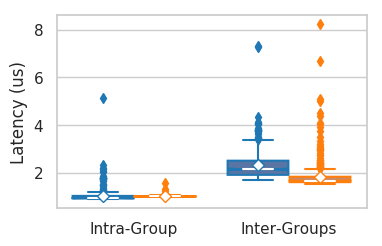}
    \caption{Latency $L$.}
    \label{fig:awr:routing:lat}
\end{subfigure}%
\hfill
\begin{subfigure}[b]{.25\textwidth}
    \centering
    \includegraphics[width=\columnwidth]{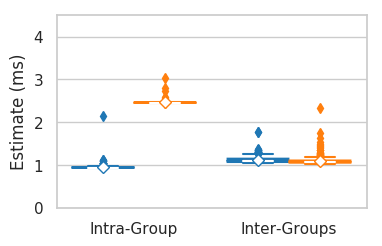}
    \caption{Time Estimation.}
    \label{fig:awr:routing:est}
\end{subfigure}%
\caption{Performance comparison of the \pingpong benchmark, when using different routing algorithms. The test has been executed on the \daint machine.}
    \label{fig:awr:routing}
\end{figure*}

To analyze the extent of this variability, we consider the performance of a \pingpong benchmark between two nodes exchanging a 4\review{MiB} message, considering some of the routing strategies described in Section~\ref{sec:background:routing}. To avoid situations where transient OS noise or network noise would affect a single routing strategy, we alternate the routing algorithm on successive \texttt{ping-pong}s. 

We report in Figure~\ref{fig:awr:routing} the performance of an iteration of the benchmark, when executed between two nodes in the same group (\textit{Intra-Group}) and when executed between two nodes in different groups (\textit{Inter-Groups}).
We show both the execution time, the latency $L$, the stalls ratio $s$, and the time estimation according to the model shown in Equation~\ref{eq:tmsg}. 
First of all, \review{whereas} for the \textit{Intra-Group} case the time required to complete a ping-pong is lower when using \adaptive routing, for the \textit{Inter-Groups} case \adaptive routing leads to an increase in network noise and to longer execution time. 

Let us start analyzing the \textit{Intra-Group} case. By analyzing the average stalled cycles $s$ (Fig.~\ref{fig:awr:routing:srout}) we observe that packets sent with \adhigh routing are affected by more stalls than \adaptive. Indeed, \review{because} the \adaptive algorithm has a higher probability of selecting a non-minimal path, on average it will distribute the packets on a wider set of paths with respect to the \adhigh algorithm, thus decreasing the average stalls per flit. \review{Because} the two routing algorithms are characterized by a similar latency (Fig.~\ref{fig:awr:routing:lat}), the stalls determine the performance difference, as also estimated by our cost model (Fig.~\ref{fig:awr:routing:est}).

On the other hand, in the \textit{Inter-Groups} scenario, given the higher number of minimal paths connecting the two nodes, \adhigh routing algorithm can better distribute the packets, and the average number of stalled cycles decreases compared to the \textit{Intra-Group} case (Fig.~\ref{fig:awr:routing:srout}). However, due to the higher number of hops between the two nodes, the latency of both routing algorithms increases (Fig.~\ref{fig:awr:routing:lat}). Moreover, \adaptive routing is characterized by higher latency variations with respect to \adhigh \review{ because}, due to \textit{phantom congestion}, sometimes it may select a non-minimal path even if that was not necessary. Indeed, if the non-minimal path was selected to avoid actual congestion, we should see the effect of congestion on \adhigh in the form of higher average latency.

As a consequence, for the \textit{Inter-Groups} case \adhigh performs better than the \adaptive routing algorithm \review{because} it is characterized by a lower latency and a comparable number of stalled cycles with respect to \adaptive.
This evaluation clearly shows that a large part of the network noise can be attributed to the \adaptive routing algorithm and that, under certain conditions, \adhigh may perform better due to lower average latency. We will show in Section~\ref{sec:evaluation}, when validating our application-aware routing algorithm, how several other microbenchmarks and real applications are also affected by the selection of the routing algorithm.
%\todo{Should we show that stalls are usually higher for adaptive with bias, while latency is usually lower? I already have the plot for that but maybe is not necessary.}

\subsection{Noise-adaptive active routing}
%- most apps execute in phases and given the previous results, different routing methods may be required for different phases ... we show a simple and effective algorithm
By leveraging these considerations, we can now devise an algorithm that uses information about latency and stalls to automatically change the routing algorithm according to the workload. As we have shown in Figure~\ref{fig:awr:routing} the optimal choice does not only depend on the workload but also on its allocation, and for this reason we cannot derive any static solution to solve this problem. Moreover, by applying a static decision it would not be possible to react to transient changes in the network conditions, such as a temporary increase in the latency \review{due to interfering jobs}, or to intrinsic changes between application phases. 
For these reasons, we rely on a runtime approach which, after a message is sent, collects counters for latency and stalls. When sending a message, the algorithm will use the counters collected for the previous message to decide which routing algorithm should be used to send the current message.

To perform this decision, we assume that the application starts by using \adaptive routing. Starting from Equation~\ref{eq:tmsg}, we denote with $L_{ad}$ and $L_{bs}$ the latencies of the \adaptive and \adhigh algorithms, respectively. Similarly, we denote the average stalls ratio with $s_{ad}$ and $s_{bs}$. Then, when sending a message \review{comprising} $f$ flits, the application would switch to \adhigh algorithm if:
\begin{equation}
    \frac{p + 512}{1024} \cdot {L_{bs}} + f\cdot(s_{bs} + 1) < \frac{p + 512}{1024} \cdot {L_{ad}} + f\cdot(s_{ad} + 1)
\end{equation}

i.e. if the message has a number of flits such that:

\begin{equation}\label{eq:awr:switch}
    f < \frac{L_{ad} - L_{bs}}{s_{bs} - s_{ad}}\cdot \frac{p + 512}{1024}
\end{equation}

\review{Because} we are assuming the application is already using the \adaptive routing algorithm, we use the $L_{ad}$ and $s_{ad}$ monitored for the last message which was sent. $L_{bs}$ and $s_{bs}$ are instead estimated by multiplying $L_{ad}$ and $s_{ad}$ by appropriate scaling factors $\lambda_{ad}$ and $\sigma_{ad}$, which we can derive by considering a median case over several runs of different microbenchmarks in different allocations. To correct mispredictions due to wrong choices of these scaling factors, the algorithm will store the last values observed for latency and stalls for both routing algorithms. These values are discarded after a given number of samples, to avoid relying on data related to a different application phase.
It is worth noting that this approach avoids the problems described in Sec.~\ref{sec:estimation:correlation} and Sec.~\ref{sec:estimation:contention}, \review{because} it only relies on NIC counters and is based on quantities which are independent from host-side delays.

\review{Because} reading the network counters for every message can introduce overhead on the application, we keep a cumulative counter of the message sizes, and we apply the algorithm when this counter is higher than a threshold (experimentally set to 4\review{KiB}). If the cumulative size is lower than the threshold, the message is sent with \adhigh routing. The reason behind that is that small messages are more affected by latency and \adhigh is usually characterized by a lower latency with respect to the \adaptive algorithm. We decided to consider the cumulative size rather than only the size of the current message to avoid that applications which always send messages smaller than the threshold would never trigger the algorithm.  
To decide when to switch from \adhigh to \adaptive, the dual equation of Equation~\ref{eq:awr:switch} can be derived.

\begin{algorithm}
\SetKwProg{selectRouting}{Function \emph{selectRouting}}{}{end}
%\SetKwProg{updateCounters}{Function \emph{updateCounters}}{}{end}
%\updateCounters{}{
%    L, s $\leftarrow$ readCounters()\;
%}

\selectRouting{msgSize}{
    \eIf{currentRouting == \adaptive}{
        $L_{ad} = L$\;
        $s_{ad} = s$\;
        \If{$L_{bs}$ and $s_{bs}$ too old}{
            $L_{bs} = L_{ad} \cdot \lambda_{ad}$\;
            $s_{bs} = s_{ad} \cdot \sigma_{ad}$\;
        }
        $f$ = getNumFlits(msgSize)\;
        \eIf{$    f < \frac{L_{ad} - L_{bs}}{s_{bs} - s_{ad}}\cdot \frac{p + 512}{1024}$}{
            currentRouting = \adhigh \;
        }{
            currentRouting = \adaptive \;
        }
    }{
        // Similar to the other branch
    }
    \Return currentRouting\;
}
\caption{Application-Aware Routing}
\label{alg:awr}
\end{algorithm}

\begin{figure*}[htpb]
\begin{center}
\begin{subfigure}[b]{.5\textwidth}
    \centering
    \includegraphics[width=\columnwidth]{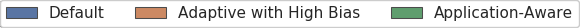}
\end{subfigure}%
\end{center}

\begin{subfigure}[b]{\textwidth}
    \centering
    \includegraphics[width=\columnwidth]{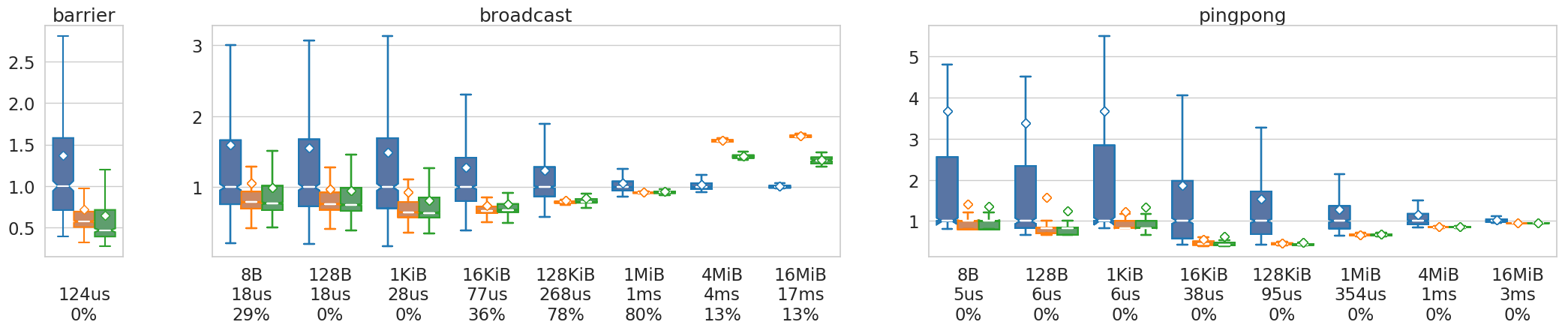}
\end{subfigure}%

\begin{subfigure}[b]{\textwidth}
    \centering
    \includegraphics[width=\columnwidth]{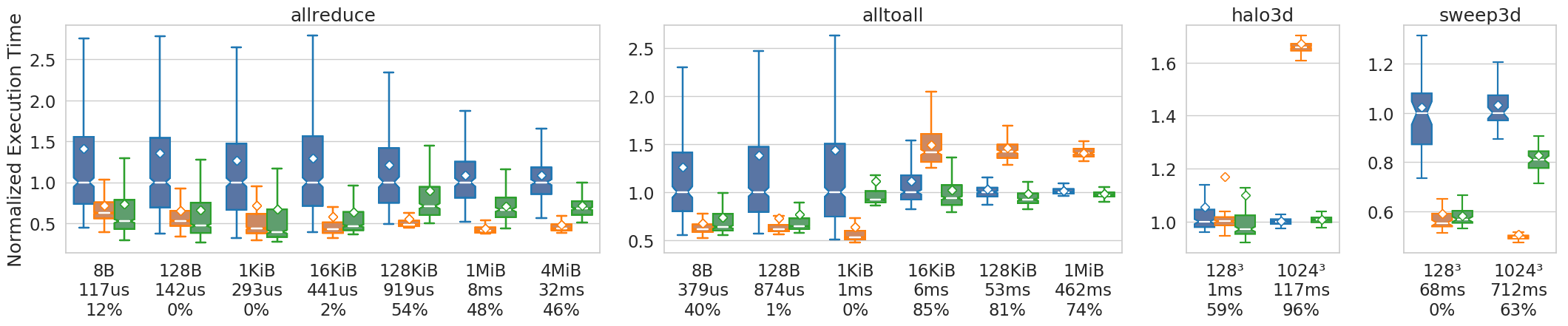}
\end{subfigure}%
\hfill

\caption{Execution time of microbenchmarks under \default, \adhigh and \awr routing algorithms, normalized with respect to the median of the \adaptive algorithm, running on 1024 nodes on \daint. For each test, we have on the x-axis: the input size, the median of the execution time for the \default routing, and the percentage of traffic which the \awr routing sends by using the \default routing.}
    \label{fig:evaluation:bench:overview:daint}
\end{figure*}

The pseudocode of this decision process is shown in Algorithm~\ref{alg:awr}. The function \textbf{\textit{selectRouting}} is called before sending the message and selects the optimal routing algorithm. After sending the message, latency and stalls network counters are read and stored into the $L$ and $s$ variable. Counters are read after sending the message to do not introduce delays in the transmission. Moreover, for \texttt{MPI\_Alltoall} communication, \adaptive is replaced with \imb routing \review{because}, as described in Section~\ref{sec:background:routing}, this is also the current default behaviour.

%\todo{make sure people get that the idea generalizes}\htor{I'd add a figure again}

\subsection{Implementation details in Aries}
\review{Because} the environment variables described in Section~\ref{sec:background:routing} can only be used to change the routing before running the application, we had to adopt a different approach for changing the routing strategy message by message. High-performance communication libraries optimized for Cray Aries network, such as Cray's \mpich, \pgas and others, rely on the \gni and \dmapp APIs, which provide low-level communication services to user-space software. Before calling any of the \gni and \dmapp functions to send messages over the Cray Aries network, it is possible to specify the routing algorithm to be used to send that message, usually by means of function parameters. However, this functionality is not exposed by higher-level APIs such as MPI. Accordingly, we implemented a dynamic library which defines the same functions used by \gni to transmit on the network (with the same signatures). Inside each of these functions, the \textbf{\textit{selectRouting}} function is called, the real \gni function is invoked by specifying the chosen routing algorithm in the call and eventually, network counters are collected by using the PAPI library~\cite{papi}. 

Every application which uses communication libraries which rely on \gni (such as MPI), can then benefit from our application-aware routing by simply specifying our library in the \texttt{LD\_PRELOAD} environment variable. A similar approach can be adopted for the \dmapp calls. \review{This allows the algorithm to be applied transparently on almost any HPC application running on a Cray Aries platform. Moreover, in principle the algorithm could also be implemented on other networks relying on adaptive non-minimal routing, if appropriate mechanisms are provided for monitoring the network state and for changing the routing algorithm at runtime.}

\section{Experimental Evaluation}\label{sec:evaluation}

In this section, we first analyze the performance of our application-aware adaptive routing on some microbenchmarks. Then, we will test it on some real applications.

\subsection{Microbenchmarks}
For the first part of our evaluation, we consider the following microbenchmarks.
\begin{description}
\item[\pingpong, \allreduce, \alltoall, \barrier, \broadcast] These benchmarks use some common MPI calls. For \pingpong, \alltoall and \broadcast the size of the messages are expressed in bytes. \texttt{Allreduce} performs a sum reduction on an array of integers and the size of the messages is expressed as the number of elements of the array.
\item[\halo] This benchmark performs a nearest neighbor communications, using a 3D stencil. Processes are logically arranged as a cube and each process communicates with the neighbors on the face of the cube formed around it as the center process. The input size corresponds to the size of the domain. For this benchmark, we used the implementation provided by the \textit{ember} benchmark suite~\cite{ember}.
\item[\sweep] This benchmark represents a wavefront communication pattern on a 3D grid. The pattern starts at a corner of the grid and \quotes{sweeps} out in a wavefront. The input size corresponds to the size of the domain. We use the implementation provided with the \textit{ember} benchmark suite~\cite{ember}.
\end{description}

To avoid situations where transient OS noise or network noise would affect a single routing strategy, for each benchmark we alternate the routing algorithm on successive iterations. \review{Because on Dragonfly networks the global available bandwidth depends on the number of nodes and links in the system, we verified that the system state (and thus the available bandwidth) did not change between different experiments.} Moreover, to reduce the impact of resources contention we only execute one process per node.

We report in Figure~\ref{fig:evaluation:bench:overview:daint} an overview of the performance of our algorithm across the microbenchmarks, for different input sizes, when executed on 1024 nodes on the \daint machine. In this experiment, the job was allocated on 257 Aries routers spanning over 6 groups. We use a fixed allocation to avoid the problems described in Section~\ref{sec:estimation:allocation}. 
We show the execution time, normalized with respect to the median of the \default routing algorithm so that values lower than 1 represent a lower execution time compared to the \default algorithm. \review{Because} outliers can be some orders of magnitude higher than the median (as we saw in Section~\ref{sec:estimation:allocation}), \review{although} we still account them for computing means and medians, we do not show them on the plot to avoid shrinking the boxes too much. On the x-axis, we report for each test the input size, the median of the execution time for the \default routing strategy and the percentage of traffic that \awr routing sends by using \default routing. It is worth mentioning that in this specific case by reasoning in terms of execution time we are not affected by the problem described in Section~\ref{sec:estimation:contention}. Indeed, if an application would be affected by OS noise or resources contention, this would affect all the routing algorithms in the same way, \review{because} the selection of the routing strategy does not have any host-side effect.

\begin{figure*}[htpb]
\begin{center}
\begin{subfigure}[b]{.5\textwidth}
    \centering
    \includegraphics[width=\columnwidth]{figs/5.1/legend.png}
\end{subfigure}%
\end{center}

\begin{subfigure}[b]{\textwidth}
    \centering
    \includegraphics[width=\columnwidth]{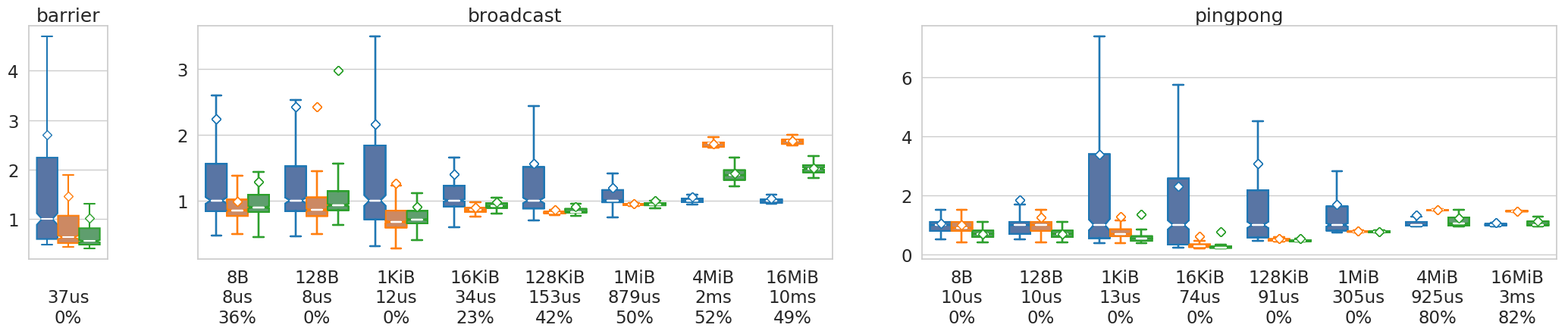}
\end{subfigure}%

\begin{subfigure}[b]{\textwidth}
    \centering
    \includegraphics[width=\columnwidth]{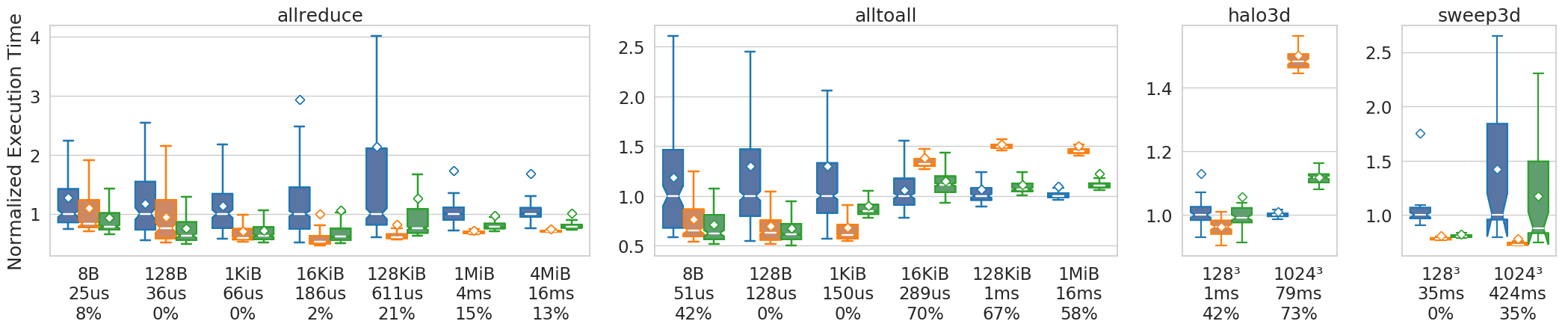}
\end{subfigure}%
\hfill

\caption{Execution time of microbenchmarks under \default, \adhigh and \awr routing algorithms, normalized with respect to the median of the \adaptive algorithm, running on 64 nodes on \cori. For each test, we have on the x-axis: the input size, the median of the execution time for the \default routing, and the percentage of traffic which the \awr routing sends by using the \default routing.}
    \label{fig:evaluation:bench:overview:cori}
\end{figure*}

\review{First}, differences between \default and \adhigh are present across different communication patterns, and in some cases, by using a different routing algorithm we can reduce the execution time by half. \review{These differences depend from the communication pattern, the amount of data exchanged between the nodes and from their allocation.} In general, \review{whereas} \adhigh performs better for benchmarks that do not generate much data (e.g., \pingpong or \barrier), when the traffic intensity is higher (e.g., for \alltoall, \broadcast and \halo), the \default routing algorithm outperforms \adhigh for large inputs. \review{Besides, for some benchmarks the gap between the different routing algorithms is larger than for others, due to the different ability to \textit{absorb} the noise. Namely, due to differences in communication and computation overlap, an increase of the average message latency may affect some benchmarks more than others.}

Moreover, when using \adhigh we have a reduction in performance variability in almost all the cases, which clearly shows that a large part of the network noise we observe is due to the choices made by the routing algorithm rather than to actual congestion on the network. To further confirm this thesis, we can observe how for each benchmark, the performance variability of \default decreases when increasing the input size, due to the lower impact of latency on the transmission time of the messages. Our \awr routing selects most of the times the algorithm which provides the best performance among the two. For example, in the \alltoall microbenchmark, it selects the \adhigh routing algorithm for small inputs \review{whereas} it selects the \default algorithm for larger message sizes.

However, despite it performs well in general, there are cases where it fails in selecting the best algorithm, such as for \broadcast of large messages and for $1024^3$ \sweep. By further investigating the issue, we found out that this is caused by oscillations in the monitored stalls and latency, i.e., as soon as the algorithm detects that the \default algorithm should be used, the stalls start to decrease and it switches back to the \adhigh algorithm and the algorithm do not converge to the best routing algorithm.
In other cases, even if \awr properly selects the optimal routing algorithm, we experience a performance drop with respect to the case where that routing strategy is statically set, such as for 1\review{KiB} \texttt{alltoall}s. This is due to the performance overhead introduced when reading the network counters and could be easily overcome by having more efficient access to network counters.
%However, we have to deal with the fact that we are trying to improve with a software-only approach, something which is done at the hardware level, and to overcome the current limitations of our \awr routing we may need some hardware support, or at least a more efficient way to access network counters.

Eventually, in some cases, \review{although} \default routing provides better performance than \adhigh, \awr can achieve the same (or even better) performance than \default even by not sending the $100\%$ of the traffic using such routing algorithm. By sending some data with \adhigh routing, less traffic is sent on non-minimal paths, reducing the network traffic. For example, this is the case of \texttt{alltoall}s larger than 16\review{KiB}. We performed the same set of experiments on 64 nodes on \cori, using 33 routers scattered on 5 Aries groups, obtaining similar results, as reported in Figure~\ref{fig:evaluation:bench:overview:cori}.

\subsection{Applications}
We now analyze our algorithm on the following applications:

\begin{description}
\review{\item[CP2K] Performs atomistic and molecular simulations of solid state, liquid, molecular, and biological systems~\cite{cp2k}}.
\review{\item[WRF-B and WRF-T] Simulations of a baroclinic wave and of a tropical cyclone performed with WRF, a weather prediction system designed for both atmospheric research and operational forecasting applications~\cite{wrf}.}

\review{\item[LAMMPS] A molecular dynamics code that models an ensemble of particles in a liquid, solid, or gaseous state~\cite{lammps}.}
\review{\item[Quantum Espresso] A suite for electronic-structure calculations and materials modeling at the nanoscale~\cite{qe}.}
\review{\item[Nekbone] Exposes the principal computation kernels of Nek5000, a fast and scalable high-order solver for computational fluid dynamics, used for many real-world
applications~\cite{nekbone,nek5000}.}
\review{\item[VPFFT] This application is an implementation of a mesoscale micromechanical materials model, which simulates the evolution of a material under deformation~\cite{vpfft}.}
\review{\item[Amber] A suite that allow users to carry out molecular dynamics simulations, particularly on biomolecules~\cite{amber}.}
\item[MILC/SU3\_RMD] The \milc benchmark represents part of a set of codes written to study quantum chromodynamics (QCD) by means of numerical simulations~\cite{milc,milc-modeling}.
%We used the \texttt{SU3\_RMD} MILC application~\cite{milc}, which performs a 4D stencil computation by using point-to-point communications among neighbors, with periodic global reductions.
\item[HPCG] Exercises computational patterns matching a wide set of applications, relying on operations like sparse triangular solvers and preconditioned conjugate gradient algorithms~\cite{hpcg}.

\review{\item[BFS and SSSP] They perform, respectively, a breadth-first search and a single-source shortest path computation on a graph. We used the Graph500 reference implementation~\cite{graph500}.}
\item[Fast Fourier Transform (FFT)] It is a computation kernel which can be found in HPC applications across multiple domains. We used the benchmark provided by the \texttt{fftw} library~\cite{fftw}.
\end{description}

\begin{figure*}[htbp]
\begin{center}
\begin{subfigure}[b]{.5\textwidth}
    \centering
    \includegraphics[width=\columnwidth]{figs/5.1/legend.png}
\end{subfigure}%
\end{center}

\begin{subfigure}[b]{\textwidth}
    \centering
    \includegraphics[width=\columnwidth]{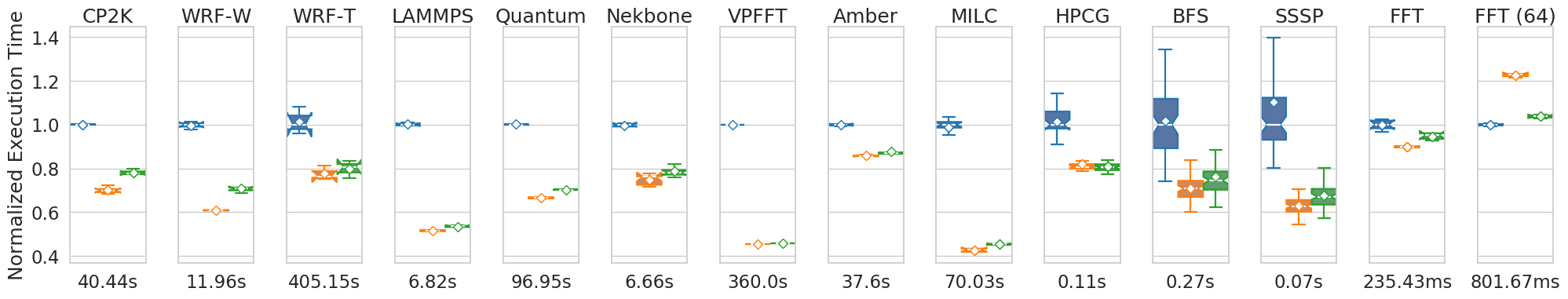}
\end{subfigure}%
\caption{Execution time of applications under \default, \adhigh and \awr routing algorithms, normalized with respect to the median of the \adaptive algorithm, running on 256 nodes on \daint.}
    \label{fig:evaluation:bench:apps}
\end{figure*}
\review{Because this work focuses on optimizing the communication phases, not all the applications can benefit from our approach, either due to their low communication intensity or to their good noise absorption capacity. For these reasons,  on such applications we did not observe any difference between the different routing algorithms, and we excluded them from our analysis~\footnote{The full results are present in the artifact.}.}
We report the results of our measurements in Fig.~\ref{fig:evaluation:bench:apps}, when executed on 256 nodes on \daint. We use in the plot the same notation used for the microbenchmarks. Even in this case, as for the microbenchmarks, a good selection of the routing algorithm can reduce the execution time of the application up to the 60\%. 

Moreover, as described in Section~\ref{sec:awr}, the best algorithm to be used does not only depend on the workload but also on the allocation and the number of nodes used to run the application. In the rightmost part of Fig.~\ref{fig:evaluation:bench:apps} we report the execution time of the \fft application when using 64 nodes. \review{While for the 256 nodes allocation \adhigh performs better than \adaptive, for the 64 nodes allocation the opposite is true.} As we can see from the results, our \awr algorithm selects the optimal routing strategy in both allocations, \review{leading to performance similar to those of \adhigh for the 256 nodes allocation and comparable to those of \default for the 64 nodes allocation.}. 
Interestingly, \review{although} both \milc and \halo use a similar communication pattern, the optimal routing strategy is different in the two cases. \review{Indeed}, \halo is a communication-oriented benchmark and does not perform any computation, \review{generating} much more traffic than \milc, thus  taking advantage of the use of \adaptive routing. 
This clearly shows how some decision which may seem optimal when stress-testing a communication network, could actually be sub-optimal for real applications. 
Overall, our \awr routing algorithm can select the optimal routing strategy across different applications and allocations, reducing the execution time by a factor of 2 on \review{several applications}.

\section{Discussion}
%\review{\paragraph{Different networks} Whereas the guidelines for isolating network noise from other types of noise can be applied to other networks, the \textit{phantom congestion} problem only arises on networks relying on adaptive routing, thus the algorithm can only be applied to such networks. Since the algorithm does not make any assumption on the network topology, it could be applied on any network using adaptive routing, independently from its topology. Concerning the implementation, uGNI and DMAPP APIs are specific to the Cray Aries platform, and they are used by the communication libraries optimized for Cray machines. We can reasonably say that the majority of HPC software running on a Cray machine would use these communication libraries either directly or by relying on some higher-level API. If appropriate mechanisms are provided for changing the routing algorithm at runtime, our algorithm could be applied to other networks relying on adaptive non-minimal routing as well.}

\review{\paragraph{Simulations} Although simulations would allow us to control different parameters and to analyze the impact and causation of noise in detail, we decided to do not rely on simulations for two main reasons. First, it is difficult to capture some aspects of the real network with a simulation, even because some information on how the network and the routing exactly work are not publicly available. Moreover, jobs are also affected by noise caused by other jobs in the system. To avoid introducing synthetically generated noise, which may not be representative of a realistic scenario, we decided to perform our analysis on a production system.}

\review{\paragraph{Static selection of the routing algorithm} Although the optimal routing algorithm depends on the size of the message to be sent and on its destination, performance differences also depend on the dynamic state of the network. For example, each rank in \halo on a $1024^3$ mesh generates, at each send, the same amount of data generated by a 64MiB \pingpong. However, even if each rank generates the same amount of data, in \halo all the ranks are communicating simultaneously, generating more traffic than \pingpong (where there are only two ranks communicating). Indeed, whereas the former benefits from using the \default routing, the latter shows better performance when using \adhigh routing. Our algorithm captures the network state by observing the average latency and the stalls monitored through network counters.}

\review{\paragraph{System state} As mentioned in Section~\ref{sec:background:routing}, traffic generated by other jobs may cross the routers used by the analyzed job. On Dragonfly networks it is not possible to isolate a job from the others because, even if we would allocate entire groups to our jobs, due to adaptive non-minimal routing packets may still traverse those groups. To mitigate these transient effects, each test has been run multiple times, and the $95\%$ confidence interval from the median (which in most cases was lower than the 5\% of the median) has been reported in the plots. Moreover, we alternated the routing algorithm on successive iterations to avoid having persistent noise affecting always the same routing algorithm. Furthermore, running the experiments on a production system without any kind of isolation allowed us to assess our algorithm in a realistic scenario where multiple and different jobs shared the network with our jobs.}

\review{\paragraph{Limitations} As shown in Section~\ref{sec:evaluation}, due to its simplicity, the algorithm has some limitations and in some cases it does not select the absolute best routing strategy. Some of these problems could be mitigated by introducing some complexity in the algorithm. However, we wanted to keep the algorithm as simple as possible because, as we shown in our evaluation, the execution of the algorithm introduces some overhead. By striving for simplicity, we kept this overhead as low as possible, both in terms of time and memory.}

\section{Related Work}
%\paragraph{OS noise} Different works analyzed the extent of noise introduced by Operating System (OS) in HPC systems, and some of them proposed solutions to mitigate such interference. ... ... \todo{to complete}

\paragraph{Network noise} Different works recently investigated the impact of network noise for different network topologies, sometimes proposing solutions to mitigate this effect. 
Some studies quantified the effect of network noise on simple communication benchmarks based on MPI collective operations~\cite{allreduce, htornnoise, Chunduri:2017:RVX:3126908.3126926, 1526010}. However, besides being affected by some of the problems we described in Section~\ref{sec:estimation}, they do not propose any solution to mitigate network noise. The work by Chunduri et al. analyzes different sources of performance variability, including a brief analysis of the impact of the routing algorithm on \texttt{MPI\_Allreduce} operations~\cite{ Chunduri:2017:RVX:3126908.3126926}. However, they do not analyze the reasons for such differences and do not exploit this information to mitigate network noise.

Most solutions optimize job allocation to minimize the contention on the links, either on dragonfly networks~\cite{ibm:tm} or on other topologies~\cite{fattree:sc18, doi:10.1142/S0129626409000419}. For example, Yang et al.~\cite{bully, 8425264} show that \review{whereas} for communication intensive jobs a random allocation is more beneficial, for less communication-intensive jobs a contiguous allocation is better. Starting from this observation, they propose a hybrid allocation scheme, to allocate communication-intensive jobs randomly \review{whereas} the less communication-intensive jobs are allocated on contiguous nodes. However, due to adaptive non-minimal routing, it is not possible to fully isolate jobs on dragonfly networks.

Bhatele et al. \cite{sc18} analyze the impact of network noise on both dragonfly and fat-tree networks, proposing an adaptive routing algorithm for fat-trees which, given the traffic matrix of the application, avoids hotspot by rerouting traffic on less loaded links. 
Eventually, some tools collect network counters across the entire network to provide visual information about congestion~\cite{tool1, tool2, overtime}.

\paragraph{Adaptive routing}
Other works analyze the limitations of adaptive routing and propose solutions to improve it.
Won et al.~\cite{7056051} shown how the \quotes{far-end} congestion should be considered as \textit{phantom} congestion, \review{because} it may be not properly represented with local information, such as the credit count. They propose a solution to overcome this problem and to avoid transient congestion, validating it by means of simulations. Other works address this problem~\cite{7161500}, by proposing  algorithms to improve congestion estimation, which are then simulated and compared to state of the art solutions. However, as also stated by the authors, approximations in the simulation are necessary \review{because} simulating the exact tiled structure of Dragonfly would be too costly.
Similarly, Faizian et al.~\cite{7801434} propose a routing scheme which also considers traffic pattern information by using network counters. After analyzing the traffic pattern, the algorithm chooses a proper bias for the adaptive routing. However, this solution relies on counters which are not available on current networks. Eventually, Jain et. al~\cite{7013015} simulate different routing strategies and their interaction with job placement.

%\todo{Is UGAL-G introduced somewhere in Sec. 2?}
%* Technology-Driven, Highly-Scalable Dragonfly Topology
%Note: 'Creation' of the dragonfly. Description of the topology and of the routing strategies. Fig 8.a. shows that MIN routing can still be better than UGAL-G.

Our main difference with respect to these works is that \review{whereas} they are usually simulated and may require changes in the hardware infrastructure, in our case we are proposing a solution which is fully software-based and does not require any modification to existing hardware and software.

%\paragraph{MISC}
% https://www.osti.gov/biblio/1368970

% https://ieeexplore.ieee.org/document/7877143

% https://www.alcf.anl.gov/files/mendygral_ALCF_ScalingDL.pdf

% https://htor.inf.ethz.ch/publications/img/widener-noise-allreduce.pdf

\section{Conclusions}
In this work we analyzed the impact of adaptive routing on network noise, proposing an application-aware routing algorithm to mitigate network noise and improve application performance. We first described how to measure variability caused by the interconnection network by isolating it from other sources of variability such as OS noise and resources contention. By following these guidelines, we shown that in some cases most of the network noise can be attributed to the adaptive routing algorithm and that noise can be reduced and performance improved by increasing the probability of selecting minimal paths.

By exploiting this knowledge we devised an application-aware routing algorithm which, before sending an application message, decides which algorithm should be used to route that message, based on information collected through network counters. Eventually, we validated this algorithm by comparing it with the \adaptive, \imb, and \adhigh algorithms provided in Cray Aries interconnection networks. We have shown how our algorithm is able to select, in most cases, the optimal routing strategy for different workloads on two different Cray machines, improving the performance up to a factor of 2 on both microbenchmarks and real applications.

%
% The acknowledgments section is defined using the "acks" environment (and NOT an unnumbered section). This ensures
% the proper identification of the section in the article metadata, and the consistent spelling of the heading.
\begin{acks}
We thank the anonymous reviewers for their insightful comments, CSCS for granting access to the Piz Daint machine, and NERSC for granting access to the Cori machine. We also thank Dr. Duncan Roweth for the useful discussions. This project has received funding from the European Research Council (ERC) under the European Union's Horizon 2020 programme (grant agreement DAPP, No. 678880). Daniele De Sensi is also supported by the EU H2020-ICT-2014-1 project \textit{RePhrase} (No. 644235) and by the University of Pisa project PRA\_2018\_66 \textit{DECLware: Declarative methodologies for designing and deploying applications}.
\end{acks}

\bibliographystyle{ACM-Reference-Format}
\bibliography{bibfile.bib}

\end{document}